\documentclass[aps,prl,twocolumn,superscriptaddress,nofootinbib,preprintnumbers,showpacs,floatfix]{revtex4-1}

\usepackage{hyperref}
\usepackage{bm}
\usepackage{amsmath}
\usepackage{graphicx}
\usepackage{psfrag}
\usepackage{subfigure}

\newcommand{\eq}[1]{Eq.~\eqref{eq:#1}}
\newcommand{\eqs}[2]{Eqs.~\eqref{eq:#1} and \eqref{eq:#2}}
\newcommand{\fig}[1]{Fig.~\ref{fig:#1}}

\newcommand{\nn}{\nonumber}
\newcommand{\sdt}{\!\cdot\!}
\newcommand{\cP}{{\mathcal P}}

\newcommand{\bfd}{{\bm d}}
\newcommand{\bfk}{{\bm k}}
\newcommand{\bfl}{{\bm \ell}}
\newcommand{\bfL}{{\bm L}}
\newcommand{\bfm}{{\bm m}}
\newcommand{\bn}{\hat{\bm n}}
\newcommand{\ord}[1]{{\mathcal O}(#1)}

\newcommand{\ff}[3]{f^{(#2)#1}_{#3}}

\newcommand{\zero}{{(0)}}
\newcommand{\one}{{(1)}}
\newcommand{\two}{{(2)}}

\newcommand{\intl}{\int\! \frac{d^2 \bfl}{(2\pi)^2}}
\newcommand{\intk}{\int\! \frac{d^2 \bfk}{(2\pi)^2}}
\newcommand{\intm}{\int\! \frac{d^2 \bfm}{(2\pi)^2}}

\begin{document}

\preprint{\vbox{
\hbox{NIKHEF 2014-003}
}}

\title{Gravitational Lensing of the CMB: a Feynman Diagram Approach}

\author{Elizabeth E.~Jenkins}
\affiliation{Department of Physics, University of California at San Diego, La Jolla, CA 92093, USA}

\author{Aneesh V.~Manohar}
\affiliation{Department of Physics, University of California at San Diego, La Jolla, CA 92093, USA}

\author{Wouter J.~Waalewijn}
\affiliation{Nikhef, Theory Group, Science Park 105, 1098 XG, Amsterdam, The Netherlands}
\affiliation{ITFA, University of Amsterdam, Science Park 904, 1018 XE, Amsterdam, The Netherlands}

\author{Amit P.~S.~Yadav}
\thanks{\mbox{Corresponding author. \href{mailto:ayadav@physics.ucsd.edu}{ayadav@physics.ucsd.edu}}}
\affiliation{Department of Physics, University of California at San Diego, La Jolla, CA 92093, USA}

\begin{abstract}
We develop a Feynman diagram approach to calculating correlations of the Cosmic Microwave Background (CMB) in the presence of distortions. As one application, we focus on CMB distortions due to gravitational lensing by Large Scale Structure (LSS).  We study the Hu-Okamoto quadratic estimator for extracting lensing from the CMB and derive the noise of the estimator up to $\ord{\phi^4}$ in the lensing potential $\phi$. By identifying the diagrams responsible for
the previously noted large $\ord{\phi^4}$ term, we conclude that the lensing expansion does not break down. The convergence can be significantly improved by a reorganization of the $\phi$ expansion. Our approach makes it simple to obtain expressions for quadratic estimators based on any CMB channel, including many previously unexplored cases. We briefly discuss other applications to cosmology of this diagrammatic approach, such as distortions of the CMB due to patchy reionization, or due to Faraday rotation from primordial axion fields.   
\end{abstract}

\maketitle

{\it Introduction\,--} 
Primary anisotropies in the Cosmic Microwave Background (CMB) were generated around $375,000$ years after the ``big bang," when the universe was still in the linear regime. The CMB field can be decomposed and studied in terms of its temperature $T$ and polarization modes $E$ and $B$. Primordial scalar perturbations create only $E$ modes of the CMB, while primordial tensor perturbations generate both parity-even $E$ modes and parity-odd $B$ polarization modes~\cite{Seljak:1996gy, Kamionkowski:1996ks, Kamionkowski:1996zd}. 
The recent detection of primordial $B$ modes~\cite{Ade:2014xna} constrains the ratio of tensor to scalar perturbations as well as the energy scale at which inflation happened~\cite{Baumann:2008aq}.

The primordial CMB generated at the surface of last scattering is statistically isotropic and Gaussian. However, during the photon's journey to us, it encounters several distorting fields, which make the CMB non-Gaussian and statistically anisotropic. Examples of such distorting fields are (a) gravitational lensing which bends the light as photons travel though the LSS~\cite{Seljak:1995ve,Zaldarriaga:1998ar,Seljak:1998nu,Zaldarriaga:2000ud}, (b) patchy reionization which modulates the CMB intensity because of scattering when Hydrogen reionizes~\cite{Dvorkin:2008tf}, and (c) cosmological rotation, due to parity-violating physics (e.g.~axions), which rotates the plane of polarization of the CMB~\cite{Carroll:1998zi,Kamionkowski:2008fp,Yadav:2009eb,Gluscevic:2009mm}.  By coupling different modes of the CMB, the distortion imprints its signature on the observed CMB by breaking statistical isotropy and introducing non-Gaussianities.
All these distortions also produce $B$-modes that contaminate the primordial tensor $B$-mode signal. 

One can utilize the statistical anisotropy of the observed CMB to reconstruct the distorting fields.  Estimators based on the Hu-Okamoto quadratic estimator~\cite{Hu:2001kj,Okamoto:2003zw} are the most studied method for extracting these distortions. 
In this paper, we present a new diagrammatic way of studying  distortions using such estimators and employ this method to investigate  the noise properties of the estimator. We  show that the previously unexplained large $N^{(2)}$ noise can be understood from the contributing diagrams, and reduced by reorganizing the expansion.
Our approach automatically yields expressions for all possible channel combinations of the quadratic estimators, including cross channels like $TE\,EB$, some of which are particularly interesting due to their low noise~\cite{Jenkins:2014hza}.

{\it Distortions in the CMB\,--} The primordial CMB is statistically isotropic and Gaussian, so all information is contained in the power spectrum, $\langle x_\bfl\, y_\bfk \rangle_\text{CMB}=C^{xy}_\bfl (2\pi)^2 \delta^2(\bfl+\bfk)$, where the average is over CMB realizations and we work in the flat sky approximation.  Here $\bfl$ and $\bfk$ denote the Fourier modes, and the power spectrum only depends on $\ell = |\bfl|$. The $x,y\in \{T,E,B\}$ are temperature and polarization components of the CMB, which can conveniently be combined into a column vector $X$ such that $C^{xy}_\bfl$ are components of a $3\times 3$ CMB power spectrum matrix $C_\bfl$,
\begin{align}
\langle X_\bfl\, X^T_\bfk \rangle_\text{CMB} &= \langle \begin{pmatrix}T_\bfl \\ E_\bfl  \\ B_\bfl\end{pmatrix} \begin{pmatrix}T_\bfk & E_\bfk &B_\bfk \end{pmatrix}\rangle_\text{CMB}
\nn \\ &
= C_\bfl (2\pi)^2 \delta^2(\bfl+\bfk)
\,.\end{align}
Secondary distortions, such as gravitational lensing, patchy reionization and Faraday rotation, will modify the components of $X$~\cite{Yadav:2009za, Dvorkin:2008tf, Yadav:2009eb,Gluscevic:2009mm, Kamionkowski:2008fp, Yadav:2012uz}. The effect of these distortion fields on Fourier modes may generically be written as
\begin{equation}\label{eq:gen}
\widetilde X_\bfl = \intm\, D_{(\bfl,\bfm)}\,X_\bfm
\,,\end{equation}
where the matrix $D_{(\bfl,\bfm)}$ can mix components of the CMB. $X_\bfm$ is the primordial spectrum, and $\widetilde X_\bfl$ is the distorted (observed) spectrum.
We now focus on how to calculate the effect of gravitational lensing using Feynman diagrams, but we will comment on other distortions in the final discussion.

{\it Gravitational lensing\,--}  
Lensing deflects the path of CMB photons from the last scattering surface. This deflection results in a remapping of the CMB temperature/polarization pattern on the sky, $\bn \to \bn + \bfd(\bn)$, and mixes the $E$ and $B$ polarization modes. The deflection is given by, $\bfd(\bn) = \nabla\phi (\bn)$,
where the lensing potential $\phi(\bn)$ is obtained by integrating the gravitational potential along the line of the sight~\cite{Seljak:1995ve}. There are higher-order corrections to the lensing potential~\cite{Bernardeau:1996un,Cooray:2002mj,Hirata:2003ka,Cooray:2005hm,Namikawa:2011cs},  which for simplicity we will ignore here.  Treating $\phi$ as a Gaussian field with power spectrum $C^{\phi\phi}_\ell$~\cite{Jenkins:2014hza},
\begin{align} \label{eq:lensing}
  D^\text{Lensing}_{(\bfl,\bfm)} &=  R_{(\bfl,\bfm)}\,(2\pi)^2 \delta^2(\bfl-\bfm-\cP)
  \nn\\ & \quad \times
 \exp\Big[ -\intk\,(\bfk \sdt \bfm)\, \phi_\bfk \Big] 
\,,\end{align}
where $\cP$ gives the total momentum of all the $\phi$ fields and $R_{(\bfl,\bfm)}$ encodes the mixing of $E$ and $B$ polarizations,
\begin{equation}
 R_{(\bfl,\bfm)} =
 \begin{pmatrix}
   1 & 0 & 0 \\
   0 & \cos 2\varphi(\bfl,\bfm) & \sin 2\varphi(\bfl,\bfm) \\
   0 & -\sin 2\varphi(\bfl,\bfm) & \cos 2\varphi(\bfl,\bfm)
 \end{pmatrix}
\,.\end{equation}
Here, $\varphi(\bfl,\bfm)$ is the (oriented) angle between $\bfl$ and $\bfm$. 
The lowest order terms in \eq{lensing} produce the familiar result~\cite{Hu:2001kj}
\begin{align}
  D^\text{Lensing}_{(\bfl,\bfm)} &= (2\pi)^2 \delta^2(\bfl-\bfm)
  -R_{(\bfl,\bfm)} \big[(\bfl-\bfm) \sdt \bfm\big]\, \phi_{\bfl-\bfm} \nn \\ & \quad+ \ord{\phi^2}
\,.\end{align}

\begin{figure}
\begin{tabular}{cl}
\raisebox{-0.5\height}{
\psfrag{labelone}[r][r]{$x$}
\psfrag{labeltwo}[l][l]{$y$}
\psfrag{labelthree}[c][c]{$\bfm$}
\includegraphics[width=4.5cm]{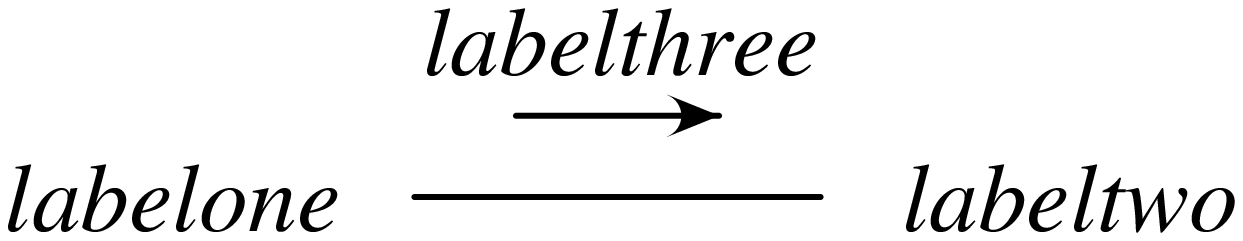}} & $C^{xy}_\bfm$ \\
\raisebox{-0.5\height}{
\psfrag{labelone}[c][c]{$\bfk$}\includegraphics[width=1.8cm]{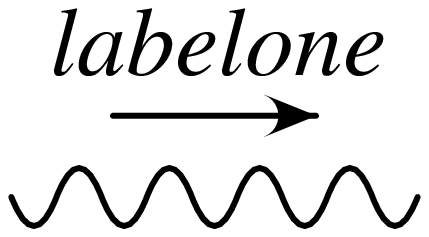}} & $C^{\phi\phi}_\bfk$ \\[1cm]
\raisebox{-0.5\height}{
\psfrag{labelone}[r][r]{$\bfl,x$}
\psfrag{labeltwo}[c][c]{$\bfm,y$}
\psfrag{labelthree}[c][c]{$\bfk_1$}
\psfrag{labelfour}[c][c]{$\bfk_2$}
\psfrag{labelfive}[c][c]{$\bfk_n$}
\includegraphics[width=5cm]{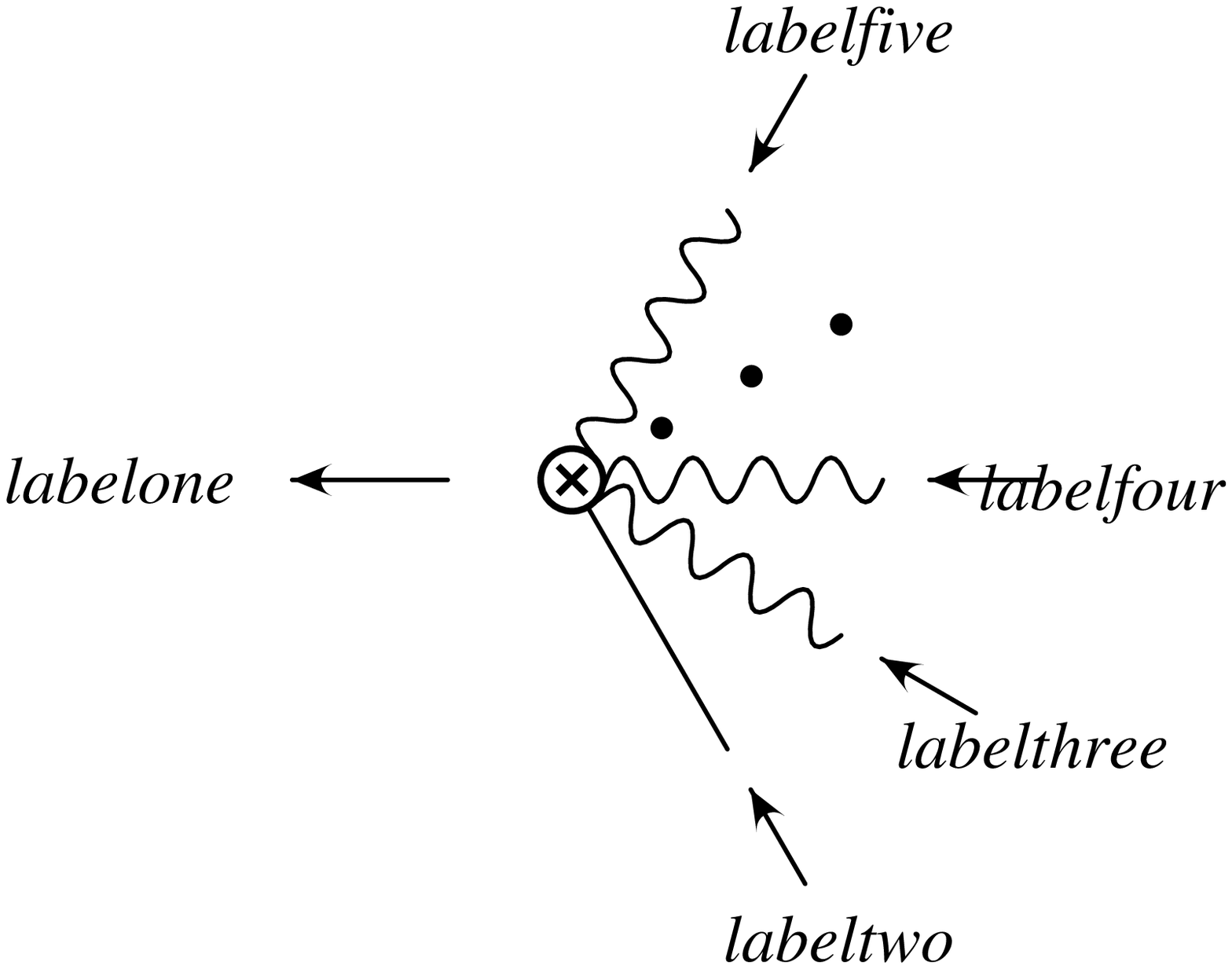}} & $\begin{array}{c} R_{(\bfl,\bfm)}^{xy} \prod_i (-\bfk_i \sdt \bfm) \\[10pt]
\text{with }\bfl = \bfm + \sum_i \bfk_i  \end{array}$ \\[2.5cm]
\raisebox{-0.5\height}{
\psfrag{labelone}[c][c]{$\bfL\ $}
\psfrag{labeltwo}[l][l]{$\bfL-\bfl,y$}
\psfrag{labelthree}[l][l]{$\bfl,x$}
\includegraphics[width=4cm]{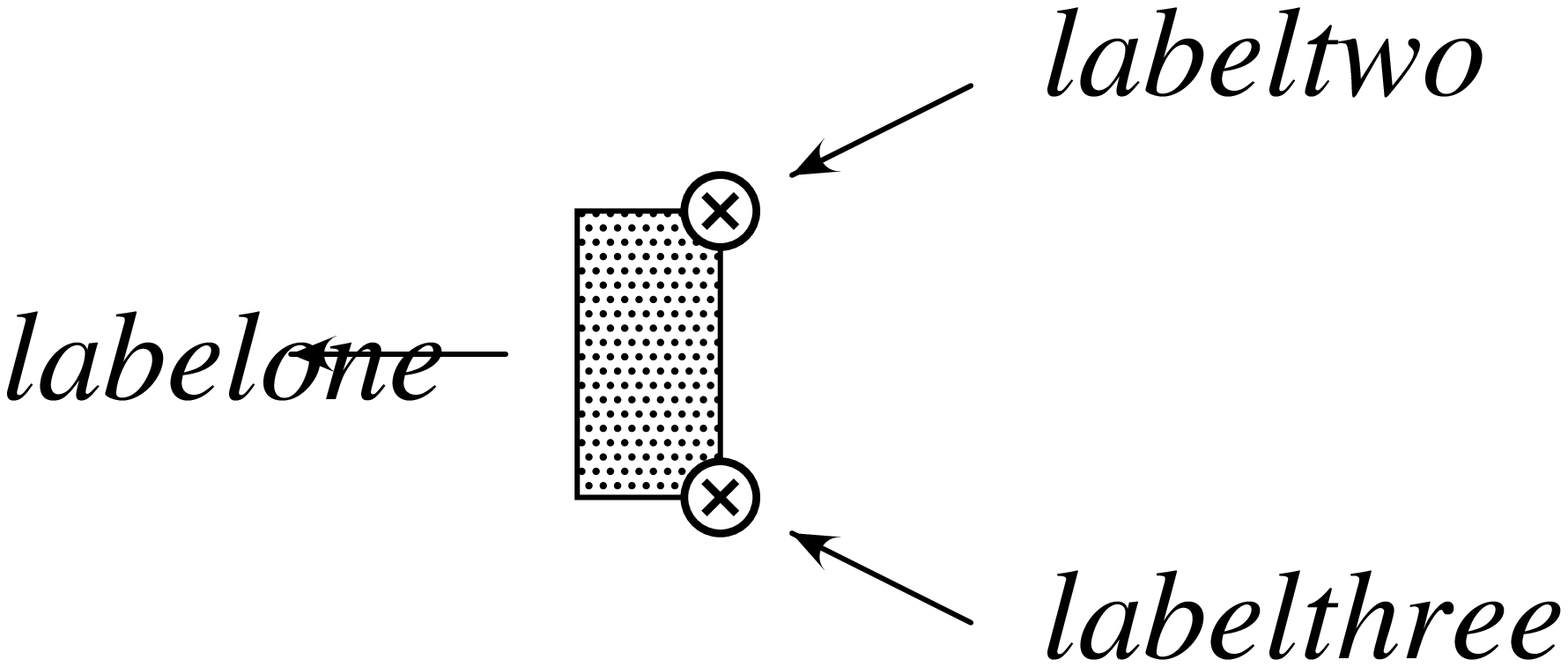}} & $\dfrac{A_{\bfL}}{L^2} F^{xy}_{(\bfl,\bfL-\bfl)}$ \\
\end{tabular}
\caption{Feynman rules for calculating lensed CMB fields.}
\label{fig:rules}
\end{figure}

{\it Feynman diagrams for lensing\,--} \eq{lensing} yields a simple Feynman rule when calculating the average of several CMB modes $\langle \widetilde x(\bfl) \widetilde y(\bfk) \dots \rangle$ over CMB or LSS realizations (see \fig{rules}). In the calculation of such an average, each lensed field $\widetilde x(\bfl)$ is represented as a vertex with momentum $\bfl$ flowing in. It has one straight line (the unlensed field) and arbitrary many wiggly lines (the lensing field $\phi$) connected to it. When averaging over CMB realizations, each straight line must begin and end at a vertex. It corresponds to $C^{xy}_\bfm$ for the CMB components $x$ and $y$, where $\bfm$ is the momentum flowing through the line. Similarly, each wiggly line corresponds to $C^{\phi\phi}_\bfk$ when averaging over LSS realizations, where $\bfk$ is the wiggly line momentum.
Momentum is conserved at the vertex and each unconstrained internal momentum $\bfk$ is integrated over with $d^2\bfk/(2\pi)^2$. In addition, there is a factor corresponding to total momentum conservation $(2\pi)^2 \delta^2(\bfl + \bfk + \dots)$ which is typically pulled out front (see e.g.~\eq{C_lensed}). These rules are summarized in \fig{rules} and will be illustrated with explicit examples below.

\begin{figure}
\begin{tabular}{ccccc}
\includegraphics[width=0.12\textwidth]{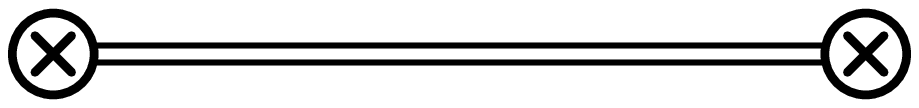} = &
\psfrag{labelone}[c][c]{$\bfl$}
\includegraphics[width=0.12\textwidth]{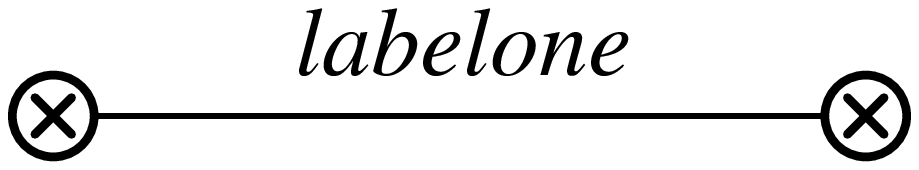} & + & 
\psfrag{labeltwo}[c][c]{\raise5pt\hbox{$k$}}
\psfrag{labelone}[c][c]{$\bfl-\bfk$}
\includegraphics[width=0.12\textwidth]{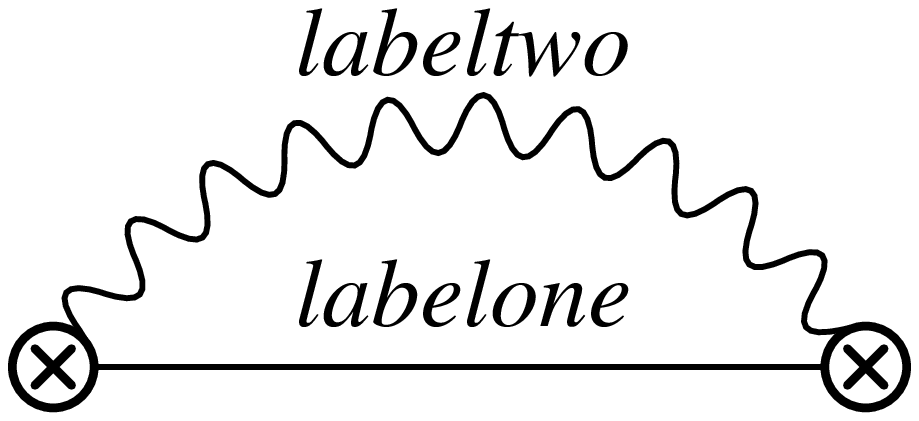} \\ 
$\widetilde C_\bfl$ & (a) & & (b) \\ &
+\psfrag{labeltwo}[c][c]{\raise5pt\hbox{$k$}}
\psfrag{labelone}[c][c]{$\bfl$}
\includegraphics[width=0.14\textwidth]{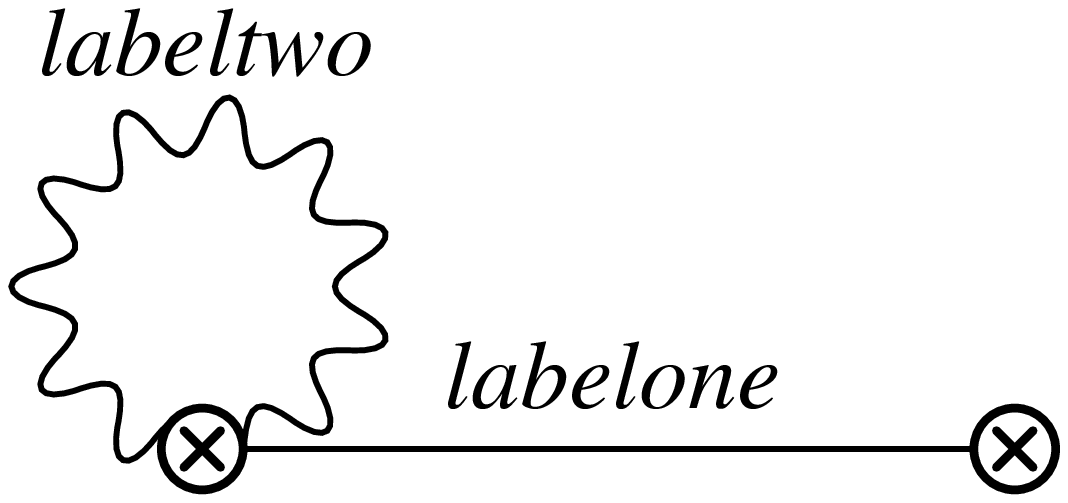} & + &
\psfrag{labeltwo}[c][c]{\raise5pt\hbox{$k$}}
\psfrag{labelone}[c][c]{$\bfl$}
\includegraphics[width=0.14\textwidth]{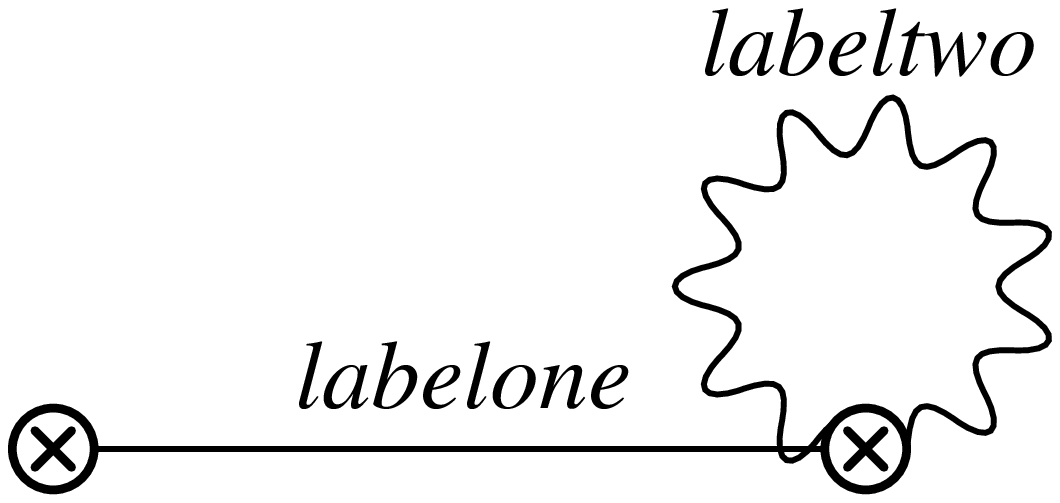} \\ &
(c) & & (d) \\
\end{tabular}
\caption{Diagrams contributing to lensed spectra (denoted by a double line). Graph (a) is the primordial contribution, and (b) - (d) describe corrections due to lensing.}
\label{fig:Cl}
\end{figure}

As a simple example, we calculate the lensed CMB spectra $\widetilde C_\bfl$. The diagrams contributing to $\widetilde C_{\bfl}$ are shown in \fig{Cl}. Using the rules from \fig{rules} gives
\begin{align} \label{eq:C_lensed}
\langle \widetilde X_\bfl \widetilde X_{\bfl'}^T \rangle  &= (2\pi)^2 \delta^2(\bfl + \bfl') \widetilde C_\bfl 
\\ &
= (2\pi)^2 \delta^2(\bfl + \bfl') \Big\{ C_\bfl +\intk  \Big [ -C_{\bfl} C^{\phi \phi}_\bfk  (\bfk\sdt \bfl)^2 
\nn \\ & \quad
+ R_{(\bfl,\bfl-\bfk)}C_{\bfl-\bfk}R^T_{(\bfl,\bfl-\bfk)} C^{\phi \phi}_\bfk \big(\bfk\sdt (\bfl-\bfk)\big)^2 \Big] \Big\}
\,,\nn\end{align} 
where both $C$ and $\widetilde C$ are $3 \times 3$ matrices in $\{T,E,B\}$.
Graph (a) is the unlensed spectrum $C_\bfl$ and graph (b) yields the third line. Graphs (c) and (d) give identical contributions, are multiplied by a symmetry factor of $1/2$, and simplify due to $R_{(\bfl, \bfl)} = 1$, giving the last term on the second line.

{\it Quadratic estimator and noise terms\,--}
Lensing breaks the statistical isotropy, correlating the CMB modes,
\begin{align}
 \langle \widetilde x_{\bfl} \widetilde y_{\bfL-\bfl}\rangle_\text{CMB} 
 &= (2\pi)^2 \delta^2(\bfL)  \widetilde C^{xy}_\bfL \! \nonumber \\
 &+ \!\big[\ff{xy}{\phi,0}{(\bfl,\bfL-\bfl)} \!+\! \ff{xy}{\phi,1}{(\bfl,\bfL-\bfl)} \!+\! \dots \big] \phi_\bfL  \nonumber\\ & \quad 
+ \intm\, \ff{xy}{\phi\phi,0}{(\bfl,\bfL-\bfl,\bfm)} \phi_{\bfL-\bfm} \phi_\bfm 
 + \dots
\,, \label{eq:2point}
\end{align}
which can be used to reconstruct the lensing field from the CMB. The superscript $0,1$ on $f$ denotes its order in powers of $C^{\phi\phi}$.
A quadratic estimator for the lensing potential can be written as
\begin{align} \label{eq:estimator}
\hat \phi^{xy}_\bfL= \frac{A^{xy}_L}{L^2} \intl\, F^{xy}_{(\bfl,\bfL-\bfl)}
\widetilde x_\bfl \widetilde y_{\bfL-\bfl} 
\,,\end{align}
and its Feynman rule is given in \fig{rules}.
Following~\cite{Hu:2001kj}, the normalization $A_L$ is chosen so that \eq{estimator} yields an unbiased estimator,  $\langle \hat \phi^{xy}_\bfL \rangle_\text{CMB} = \phi_\bfL$, and 
the filter $F^{xy}$ is determined by minimizing the variance $\langle \langle \hat \phi_\bfL^{xy} \hat \phi_{\bfL'}^{x'y'} \rangle_\text{CMB} -  \langle \hat \phi_\bfL^{xy} \rangle_\text{CMB} \langle \hat \phi_{\bfL'}^{x'y'} \rangle_\text{CMB} \rangle_\text{LSS}$ at lowest order in the lensing expansion.

\setlength{\tabcolsep}{9pt}
\begin{figure}
\begin{tabular}{cc}
\raisebox{-0.5\height}{\includegraphics[width=0.15\textwidth]{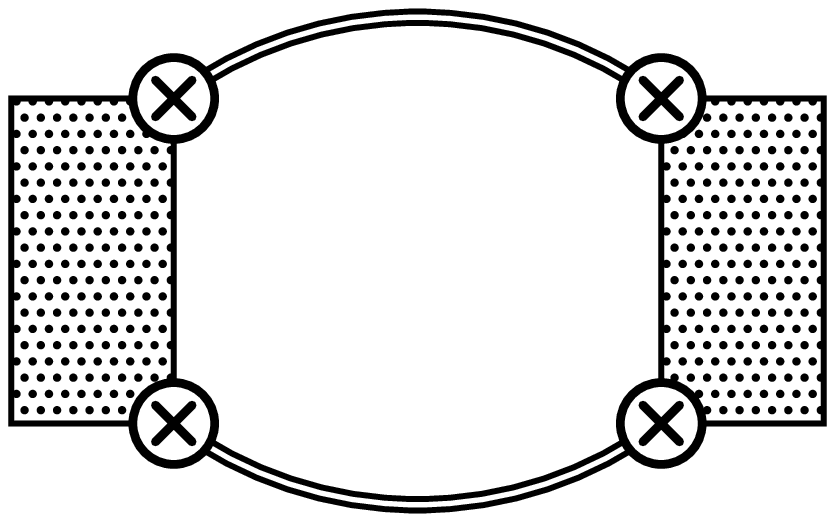}} &
\raisebox{-0.5\height}{\includegraphics[width=0.15\textwidth]{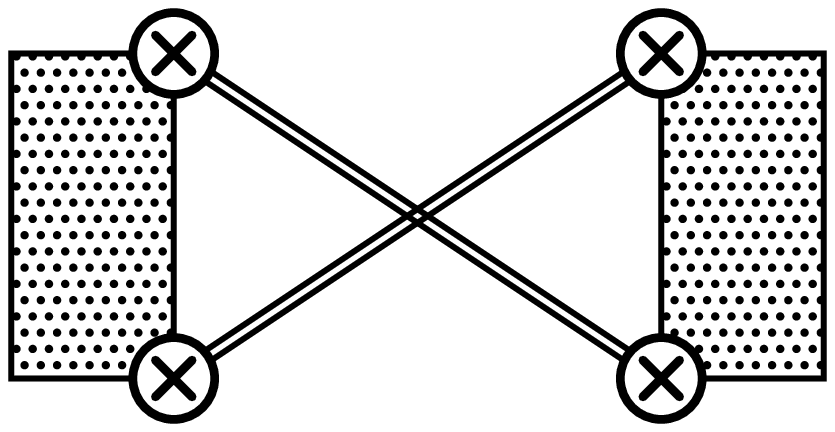}} \\ \ \\[-2ex]
(a) & (b)
\end{tabular}
\caption{Diagrams describing the lowest-order noise $N^{xy\zero}_\bfL$.}
\label{fig:N0}
\vspace{2ex}
\raisebox{-0.5\height}{
\psfrag{labelone}[c][c]{\raisebox{2mm}{$x,\bfl$}}
\psfrag{labeltwo}[c][c]{\raisebox{2mm}{$y,\bfl^\prime$}}
\psfrag{labelthree}{$\bfl+\bfl^\prime$}
\includegraphics[width=3cm]{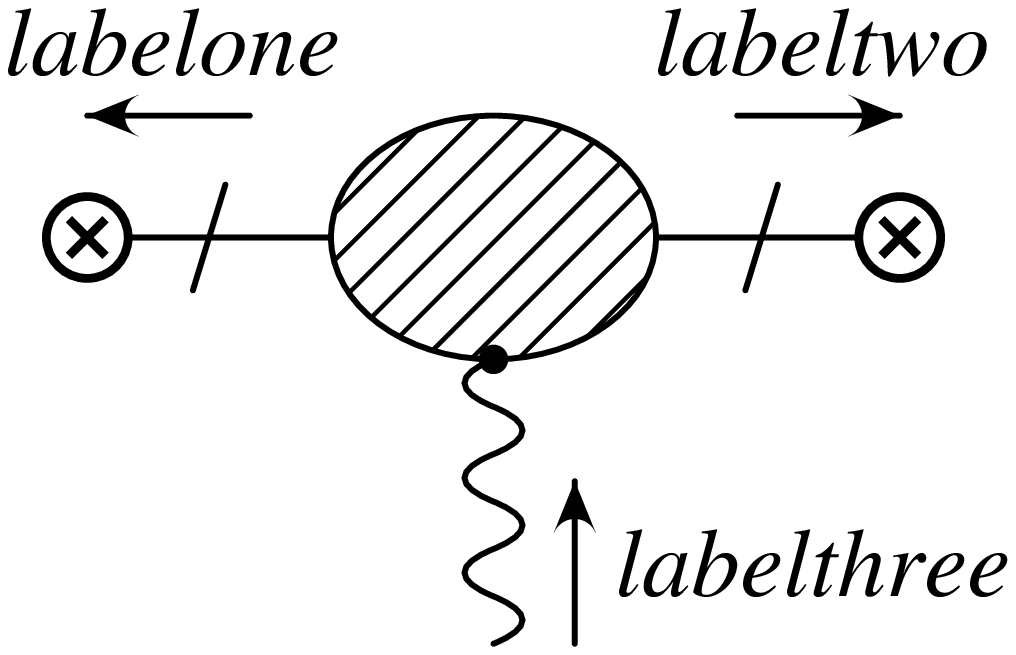}} $\!\!\!\!=$
\raisebox{-0.5\height}{\includegraphics[width=0.11\textwidth]{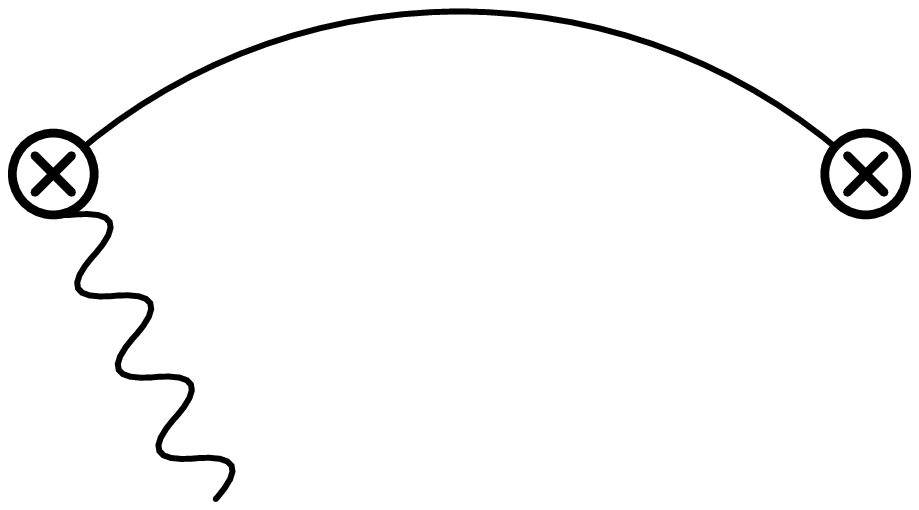}} $+$
\raisebox{-0.5\height}{\includegraphics[width=0.11\textwidth]{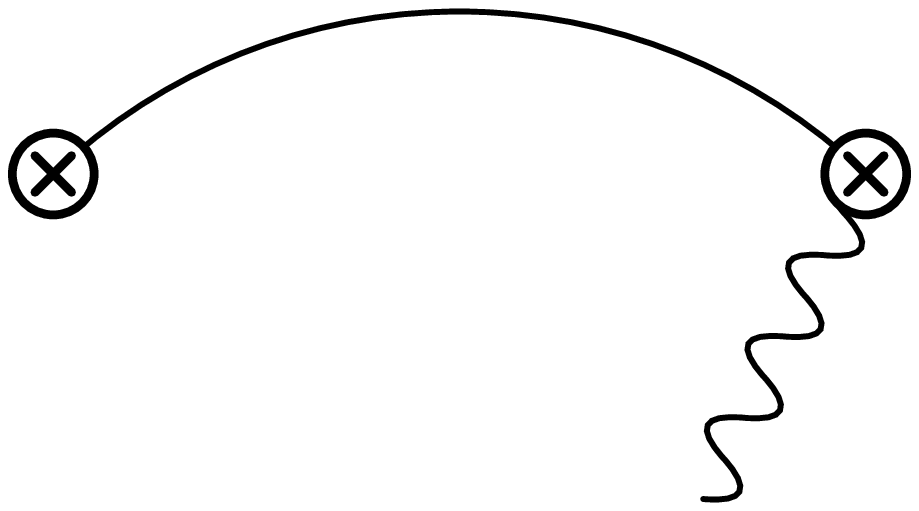}}
\caption{Diagrams contributing to the filter $f^{(\phi,0)xy}_{(\bfl,\bfl')}$.}
\label{fig:f0}
\vspace{2ex}
\begin{tabular}{cc}
\raisebox{0.42\height}{\includegraphics[width=0.2\textwidth]{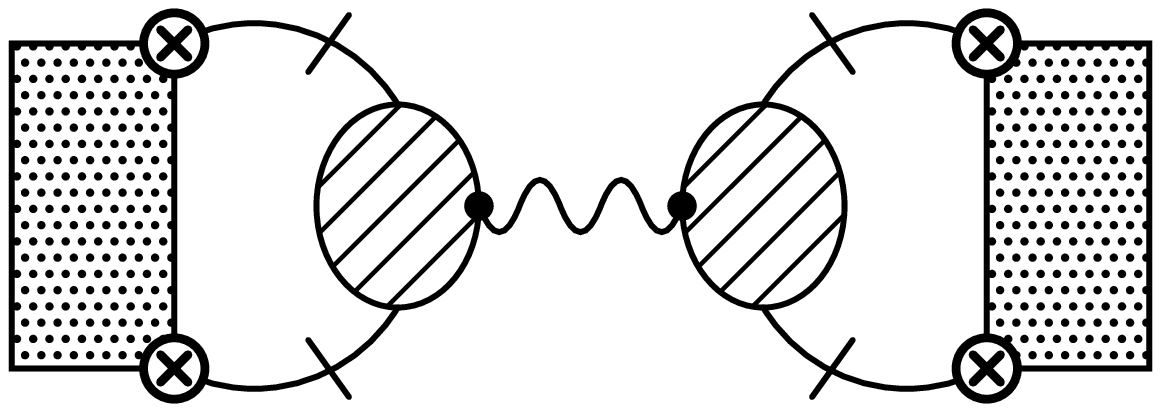}} &
\includegraphics[width=0.175\textwidth]{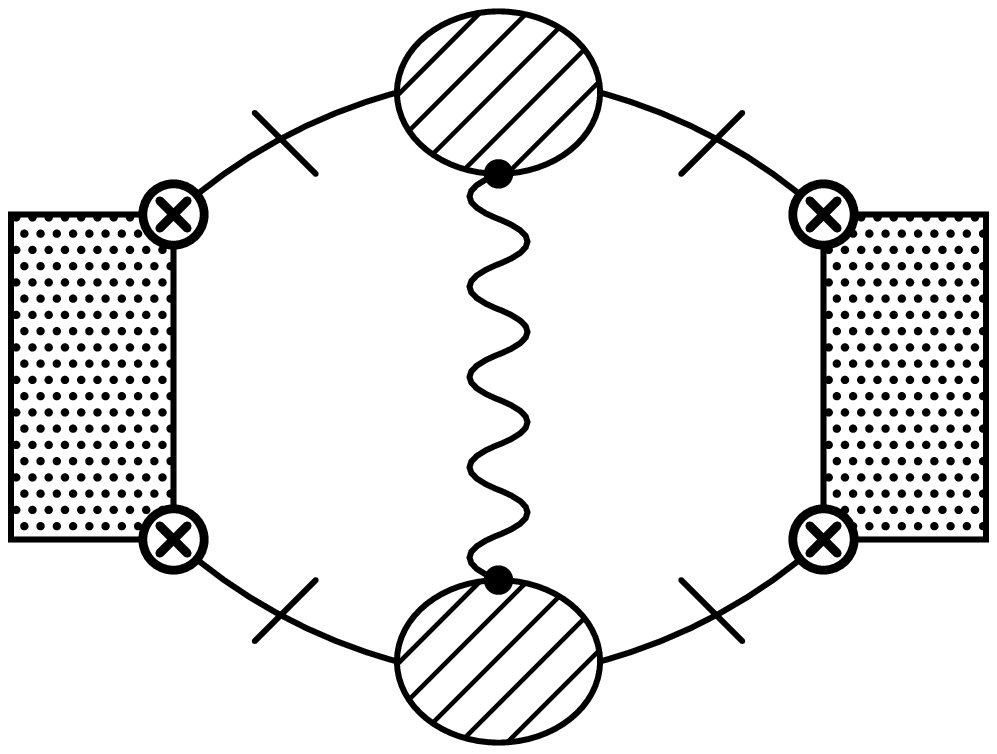} \\ \ \\[-2ex]
(a) & (b)
\end{tabular}
\caption{Diagrams contributing at $\ord{\phi^2}$. Diagram (a) produces $C^{\phi\phi}$ and (b) gives $N^\one$. The cross graph for (b) (analogous to \fig{N0}(b)) is not shown.}
\label{fig:N1}
\end{figure}

Using the quadratic estimator to extract the lensing power spectrum introduces a bias
\begin{align} \label{eq:bias}
 \langle \hat \phi_\bfL^{xy} \hat \phi_{\bfL'}^{x'y'} \rangle_\text{CMB,LSS}
&=  (2\pi)^2 \delta^2(\bfL \!+\! \bfL') \big[C^{\phi\phi}_\bfL \!+\! N^{xy,x'y' \zero}_\bfL\! 
\nn \\ & \quad
+\!N^{xy,x'y' \one}_\bfL
+\!N^{xy,x'y' \two}_\bfL \!+\! \ord{\phi^6}\big]
\,,\end{align}
given by the noise terms $N^{xy,x'y' (n)}_\bfL$, which are $\ord{\phi^{2n}}$. The Gaussian noise $N^{\zero}_\bfL$ is expected to provide the dominant contribution to the variance. However, it has been recently noticed that the higher order noise term $N^{\two}_\bfL$ can give a large contribution at small $L$~\cite{Hanson:2010rp,Anderes:2013jw}. One of the main goals of our paper is to illustrate the power of Feynman diagrams in calculating the higher-order contributions $N^{\one}_\bfL$ and $N^{\two}_\bfL$, which also makes it easy to track down the origin of this large contribution.

The two diagrams contributing to the lowest order noise term are shown in \fig{N0} and lead to
\begin{align} \label{eq:N0}
\hspace{-2ex}N^{xy,x'y'(0)}_\bfL&= \frac{A^{xy}_\bfL A^{x'y'}_\bfL}{L^4} \intl\, F^{xy}_{(\bfl,\bfL-\bfl)}
\nonumber \\
& \times \Big[  F^{x'y'}_{(-\bfl,\bfl-\bfL)} 
  {\overline C}^{xx'}_\bfl {\overline C}^{yy'}_{\bfL-\bfl} + F^{x'y'}_{(\bfl-\bfL,-\bfl)}  {\overline C}^{xy'}_\bfl {\overline C}^{x'y}_{\bfL-\bfl} \Big]
\,,\end{align}
in agreement with Ref.~\cite{Hu:2001kj}. Here ${\overline C}^{xy}_\bfl=\widetilde C^{xy}_\bfl+ \Delta^2_{xy} \, e^{\ell (\ell+1)\sigma^2/8\ln2}$ is the observed spectra,  $\sigma$ is the full-width-half-maximum of the experimental beam, and  $\Delta_{xy}$ is experimental noise~\cite{Knox:1995dq}. We will assume fully polarized detectors for which $\Delta_{EE}=\Delta_{BB}=\sqrt{2}\Delta_{TT}$, and $\Delta_{xx'}=0$ for $x \neq x^\prime$. 

\begin{figure}
\vspace{2ex}
\includegraphics[bb=18 174 572 658,width=0.5\textwidth]{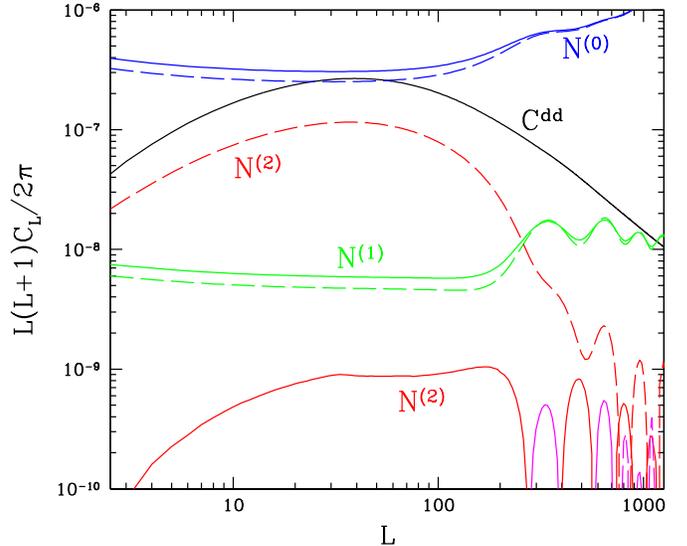}
\caption{Lensing signal $C^{dd}_L=L^2 C^{\phi\phi}_L$ and noise power spectra $N^{TT(n)}_L$ of the quadratic estimator for $C^{dd}_L$.  Lensed spectra are used for the solid curves, as described in the text, improving the convergence. We have assumed experimental noise $\Delta_{EE}=56\,\mu \text{K}$-arcmin and beam $\sigma =7'$. The sign of the noise $N^{(2)}_L$ changes as a function of $L$; negative values are shown in red and positive values in magenta.}
\label{fig:Nl}
\end{figure}

Although using lensed rather than unlensed spectra in $\overline C$ is formally beyond the order in $\phi$ of $N^\zero_\bfL$, it reduces the number of diagrams contributing to the higher-order noise. Specifically, corrections of the type shown in \fig{Cl}(b) - (d) are now already included.
This approach is standard for the Gaussian noise $N^\zero_\bfL$, but we find that also using lensed spectra in the higher-order noise terms improves their convergence. We will compare using lensed vs.~unlensed spectra when we present numerical results in \fig{Nl}. We also use lensed spectra everywhere in the filter $F$ of the estimator, which has been considered in Refs.~\cite{Lewis:2011fk,Anderes:2013jw}.

The diagrams contributing at $\ord{\phi^2}$ are shown in \fig{N1}, which we break into subgraphs involving the filter $f^{(\phi,0)}$ defined in \eq{2point}. This filter $f^{(\phi,0)}$ describes the distortion of the two-point function due to lensing, as shown in \fig{f0}. In these figures the ``crossing out" of lines indicates that they do not produce a power spectrum in the corresponding expression. \fig{N1}(a) produces the lensing spectrum $C^{\phi\phi}_\bfL$ by construction. \fig{N1}(b) and the corresponding cross graph can be calculated using the Feynman rules in \fig{rules}, 
\begin{align} \label{eq:N1}
\ff{}{\phi,0}{(\bfl,\bfl')}&=R_{(\bfl,\bfl')} C_{\bfl'}  (\bfl+\bfl')\cdot \bfl' 
\\ & \quad
+  C_\bfl R^T_{(\bfl',\bfl)} (\bfl+\bfl')\cdot \bfl 
\,,\nn\\
N^{xy,x'y'(1)}_\bfL&= \frac{A^{xy}_\bfL A^{x'y'}_\bfL}{L^4} \intl \frac{d^2 \bfk} {(2\pi)^2}\, C^{\phi \phi}_\bfk F^{xy}_{(\bfl,\bfL-\bfl)} 
\nn \\ & \quad \times
\Big[ F^{x'y'}_{(\bfk-\bfl,\bfl-\bfL-\bfk)} \ff{xx'}{\phi,0}{(\bfl,\bfk-\bfl)} \ff{yy'}{\phi,0}{(\bfL-\bfl,\bfl-\bfL-\bfk)} 
\nn \\ & \qquad
+ F^{x'y'}_{(\bfl-\bfL-\bfk,\bfk-\bfl)} \ff{xy'}{\phi,0}{(\bfl,\bfk-\bfl)} \ff{yx'}{\phi,0}{(\bfL-\bfl,\bfl-\bfL-\bfk)} \Big] 
\,.\nn\end{align}
This noise contribution was first determined by Kesden et al.~\cite{Cooray:2002py,Kesden:2003cc} (for $x=x'$ and $y=y'$). Note that corrections of the form shown in \fig{Cl}(b) through (d) were already part of the calculation of $N^\zero_\bfL$ by using lensed spectra there, and thus should not be included in $N^\one_\bfL$. We will also consider using lensed spectra $\widetilde C_\bfl$ and $\widetilde C_{\bfl'}$ instead of $C_\bfl$ and $C_{\bfl'}$ in $\ff{}{\phi,0}{}$.

\begin{figure}
\begin{tabular}{cc}
\includegraphics[width=0.175\textwidth]{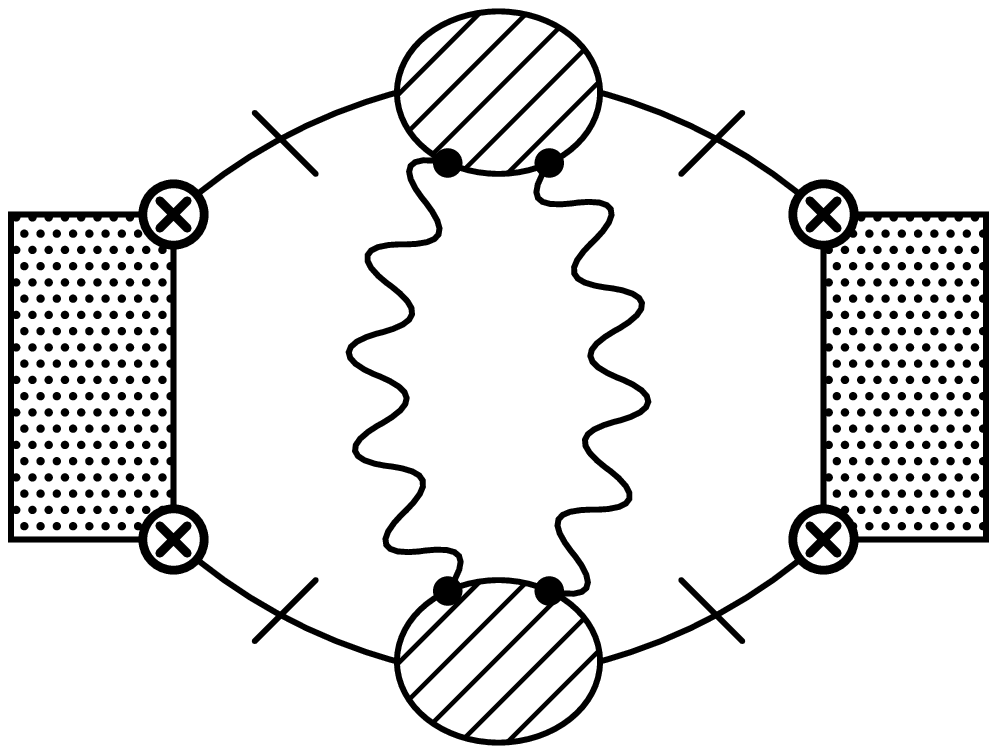} &
\includegraphics[width=0.175\textwidth]{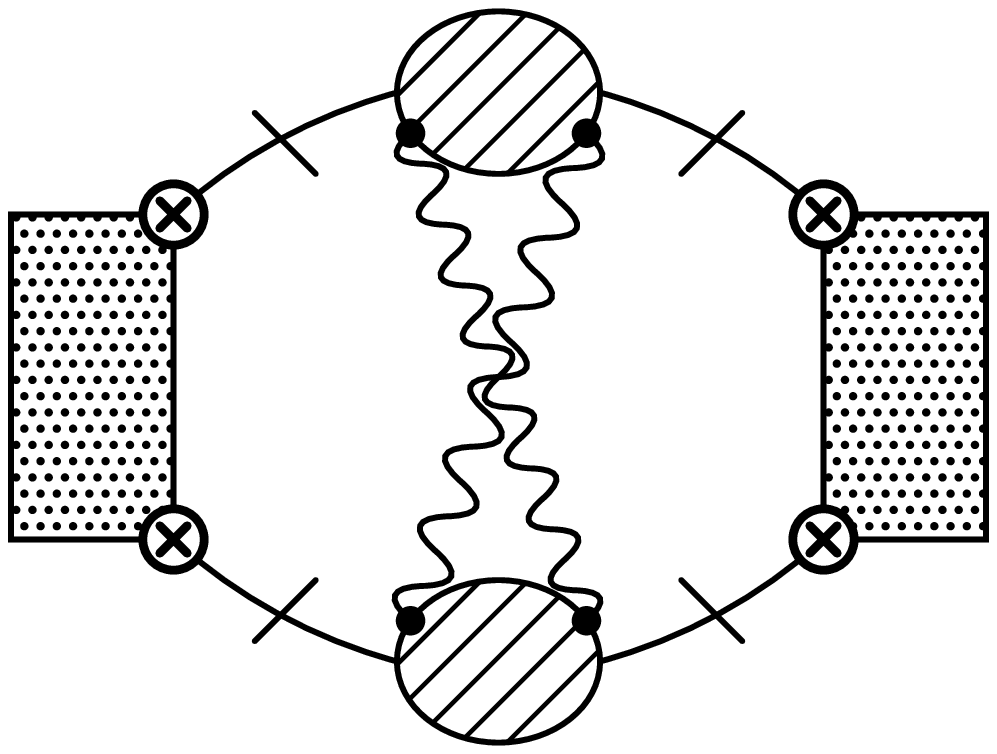} \\
(a) & (b) \\
\raisebox{-0.5\height}{\includegraphics[width=0.2\textwidth]{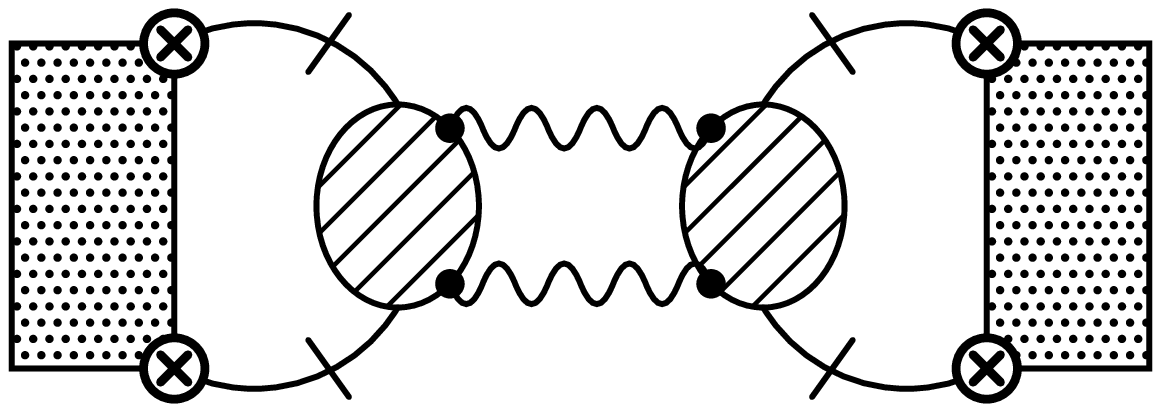}} & 
\raisebox{-0.5\height}{\includegraphics[width=0.2\textwidth]{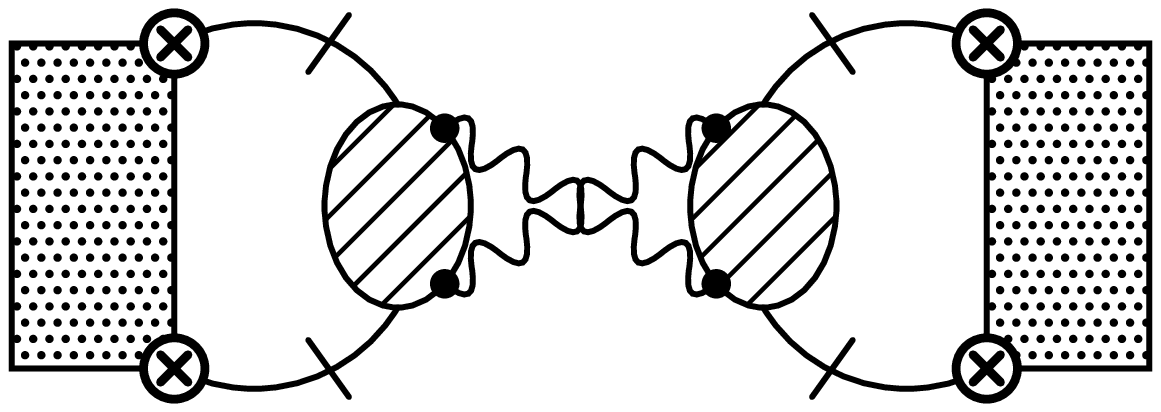}} \\
(c) & (d) 
\end{tabular}
\caption{Diagrams contributing to $N^\two$ involving $f^{(\phi\phi,0)}$. The cross graphs (analogous to \fig{N0}(b)) are not shown.}
\label{fig:N2}
\vspace{2ex}
\raisebox{-0.5\height}{\includegraphics[width=0.11\textwidth]{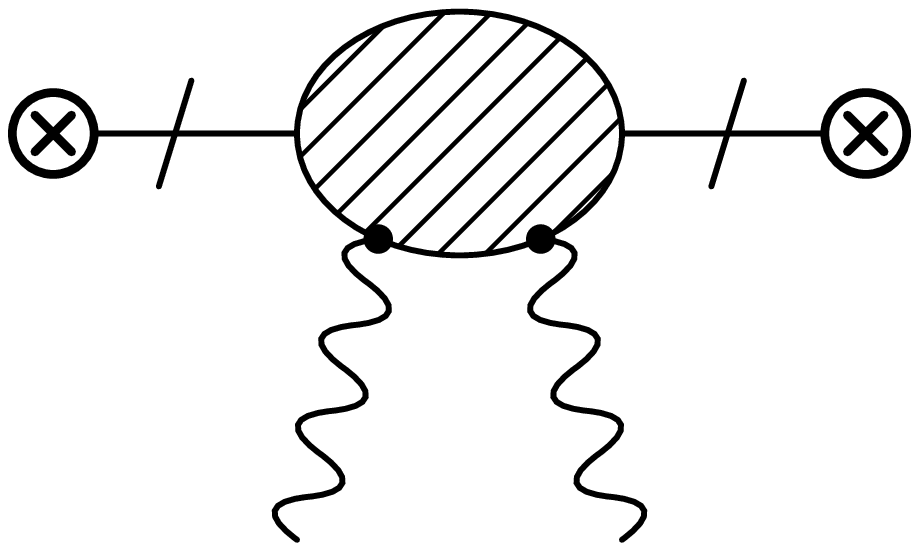}} $=$
\raisebox{-0.5\height}{\includegraphics[width=0.1\textwidth]{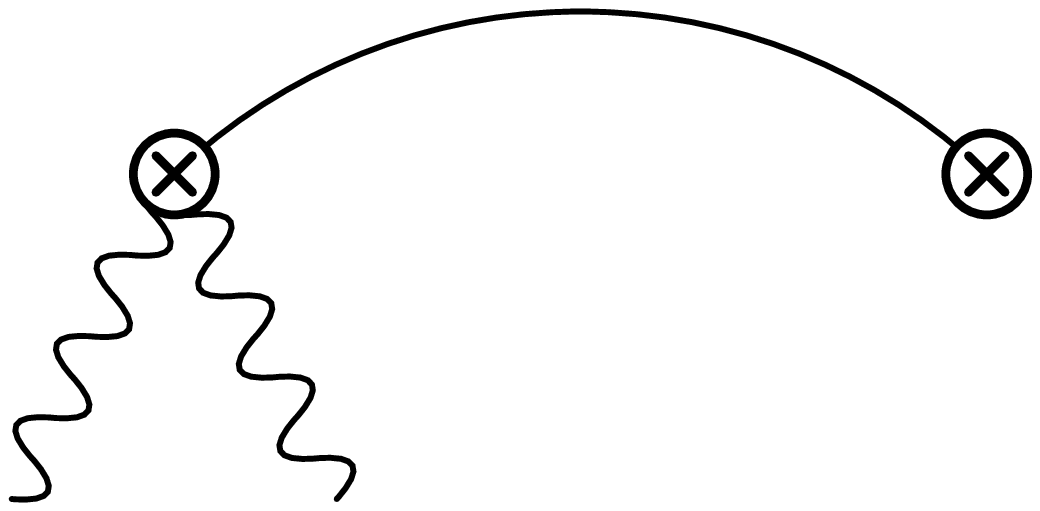}} $+$\!
\raisebox{-0.5\height}{\includegraphics[width=0.1\textwidth]{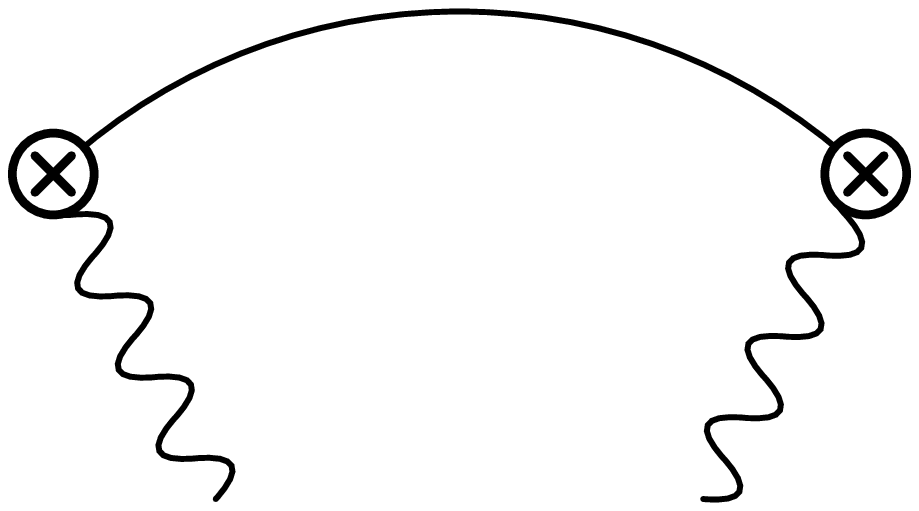}} $+$
\raisebox{-0.5\height}{\includegraphics[width=0.1\textwidth]{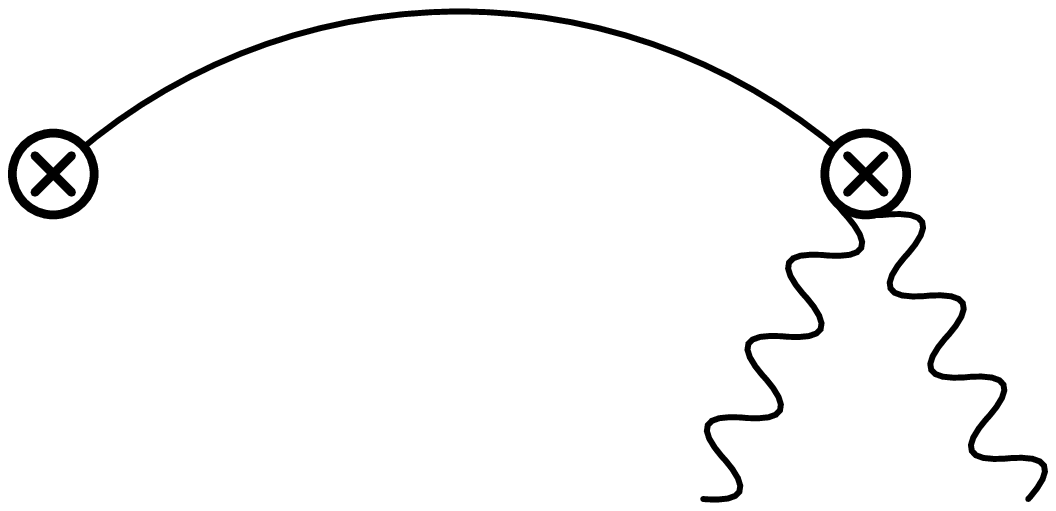}}
\caption{Diagrams contributing to the filter $f^{(\phi\phi,0)}$.}
\label{fig:g}
\end{figure}

There are two classes of diagrams contributing to $N^{(2)}_\bfL$. The first class of diagrams is the same form as those in \fig{N1} and \eq{N1}, but with one of the $f^{(\phi,0)}$ vertices replaced by the higher order $f^{(\phi,1)}$. The second class of diagrams is shown in \fig{N2}, and involves the new filter $f^{(\phi\phi,0)}$ in \fig{g}. 
Expressions for $f^{(\phi,1)}$, $f^{(\phi\phi,0)}$ and $N^{(2)}$ will be given in Ref.~\cite{Jenkins:2014hza}.
We have identified the $\ord{\phi^4}$ analogue of \fig{N1}(a) as the contribution that is responsible for the large size of $N^\two$ that had been observed in Refs.~\cite{Hanson:2010rp,Anderes:2013jw}. To understand this, it is useful to first discuss the contributions at order $\phi^2$:  \fig{N1}(a) and (b) yield $C^{\phi\phi}$ and $N^{(1)}$ and although they are formally of the same order in the lensing expansion, $C^{\phi\phi}$ is numerically larger. This is to be expected because \fig{N1}(a) has less loop integrals than (b). The same is true at $\ord{\phi^4}$ and there is thus no breakdown of perturbation theory. Instead, the graph in \fig{N1}(a) and corresponding higher order contributions give a convergent expansion but one that is numerically larger than the diagrams in \fig{N1}(b), \fig{N2}, etc.

The numerical size of this contribution can be significantly reduced by organizing the expansion in terms of lensed spectra $\widetilde C^{xy}$, i.e.~replacing $C^{xy} \to \widetilde C^{xy}$ in $f^{(\phi,0)}$ and compensating for this change in $f^{(\phi,1)}$ (with $f^{(\phi,1)}$ also written in terms of $\widetilde C^{xy}$). This essentially sums a class of higher order corrections, as we already discussed for $N^\zero$.
The results are shown in \fig{Nl}, which compares using lensed to unlensed spectra in the computation of $N^{(2)}$ and will be discussed below. Since the estimator minimizes the leading order variance, and this reorganization changes what is called leading order, the estimator is modified as well.

{\it Numerical Results\,--} In \fig{Nl}, we show the noise $N^{TT}_L$ in estimating the lensing $C^{dd}_L= L^2 C^{\phi\phi}_L$ as a function of $L$ for a Planck-like experiment with experimental noise $\Delta_{EE}=56\, \mu \text{K}$-arcmin and beam size $\sigma=7$ arcmin. We consider noise calculated using two counting methods, the unlensed spectra (dotted curves) as well as lensed spectra (solid curves). Our results for the former agree with Ref.~\cite{Hanson:2010rp}, showing that at small $L$ the bias $N^{TT\two}_L$ is large. As we explained, this originates from higher-order corrections to \fig{N1}(a), so it is not surprising that its shape is similar to $C^{dd}_L$. \fig{Nl} clearly illustrates that using lensed spectra greatly improves the convergence of the noise terms. 
The main difference with Ref.~\cite{Anderes:2013jw} is that in addition to changing our estimator to use lensed spectra, we have also reorganized our noise in terms of lensed spectra. This use of lensed spectra modifies $N^\zero$ and $N^\one$ and is responsible for the improved convergence we see, in contrast to the seemingly accidental cancellation between $N^{TT\one}_L$ and $N^{TT\two}_L$ found in Ref.~\cite{Anderes:2013jw}. 

{\it Discussion\,--} We have shown how Feynman diagrams can be used to understand the CMB, illustrating their power in the context of gravitational lensing. This method allowed us to simultaneously obtain expressions for quadratic estimators based on any CMB channel and identify the origin of the (supposed) poor convergence of higher order noise terms. Additional details, as well as plots for the polarization channels, such as $TT\, EE$ and  $EE\,EB$ are given in a subsequent publication~\cite{Jenkins:2014hza}.

Apart from lensing, there are other cosmological effects that can couple the modes of the CMB~\cite{Yadav:2009za} such as screening from patchy reionization~\cite{Dvorkin:2008tf}, and rotation of the plane of polarization either due to primordial magnetic fields~\cite{Kosowsky:1996yc,Kosowsky:2004zh, Yadav:2012uz} or parity-violating physics~\cite{Kamionkowski:2008fp,Yadav:2009eb,Gluscevic:2009mm}. The formalism presented here can be used to study these effects as well (see \eq{gen}). Below we discuss cosmological rotation and patchy reionization and show how Feynman rules can be derived for them. \\

{\it Cosmological rotation and patchy reionization\,--}
Many theories predict parity-violating primordial fields such as axions, which have  Chern-Simons couplings of the form $a F_{\mu \nu}\tilde F^{\mu \nu}$~\cite{Carroll:1998zi,Pospelov:2008gg}, that rotate the plane of polarization of light through an angle
$d\alpha = 2 \it d\tau \dot a$ during propagation for a conformal time 
$d\tau$. The fluctuations in the axion field $a$ then will be imprinted in the rotation angle $\alpha$ of the polarization. 
The observed (rotated) and primordial CMB in terms of Stokes parameters are related by $(\widetilde Q\pm i \widetilde U)({\bf n})=e^{\pm 2i \alpha(\bf n)}( Q\pm i U)({\bf n})$,
which we can write in terms of \eq{gen} as
\begin{align} \label{eq:daxion}
  D^\text{Rotation}_{(\bfl,\bfm)} &= (2\pi)^2 \delta^2(\bfl-\bfm-\cP)\, R_{(\bfl,\bfm)}
   \\ & \quad \times
  \exp\Big[ 2\lambda \intk\, \alpha_\bfk \Big] 
 \nn \\ 
  & = (2\pi)^2 \delta^2(\bfl-\bfm) 
  + 2 R_{(\bfl,\bfm)} \lambda\, \alpha_{\bfl-\bfm}
 + \ord{\alpha^2}
\,,\nn\end{align}
which is frequency independent, and mixes $E$ and $B$ through $\lambda$,
\begin{align}
  \lambda = 
  \begin{pmatrix}
   0 & 0 & 0 \\
   0 & 0 & 1 \\
   0 & -1 & 0
 \end{pmatrix}
\,.\end{align}

Reionization marks the time after decoupling when the vast majority of Hydrogen became ionized due to gravitational nonlinearities. When and how this process occurred is at present not well constrained. Inhomogeneous reionization produces several secondary anisotropies in the CMB. The patchy nature of reionization results in a  Thomson scattering optical depth to recombination, $\tau(\bf{n})$, depending on direction $\bf{n}$. Such optical depth fluctuations act as a modulation effect on CMB fields by suppressing the primordial anisotropies with a factor of $e^{-\tau(\bf{n})}$, correlating different modes by
\begin{align} \label{eq:dpatchy}
  D^\text{Reionization}_{(\bfl,\bfm)} &= (2\pi)^2 \delta^2(\bfl-\bfm-\cP)\, R_{(\bfl,\bfm)}
   \\ & \quad \times
  \exp\Big[ - \intk\, \tau_\bfk \Big] 
 \nn \\ 
  & = (2\pi)^2 \delta^2(\bfl-\bfm) 
  - R_{(\bfl,\bfm)} \tau_{\bfl-\bfm}
 + \ord{\tau^2}
\,.\nn\end{align}
From \eqs{daxion}{dpatchy}, one can obtain the corresponding Feynman rules that allow one to calculate their effect of the correlation structure of the CMB and construct the appropriate estimators and noise terms.  Assuming these effects are Gaussian and statistically isotropic, the only other ingredient is $C^{\alpha\alpha}$ and $C^{\tau\tau}$. In this case, the calculation is an expansion in $\alpha$ or $\tau$ instead of $\phi$.

{\it Acknowledgements\,--} 
The computational resources required for this work were accessed via the Glidein-WMS \cite{Sfiligoi2009} on the Open Science Grid~\citep{Pordes2007}. Numerical integrations were carried out using the {\sc Cuba} integration library~\cite{Hahn:2004fe}. 

APSY would like to thank Matias Zaldarriaga for discussions at an early stage of this project. This work was supported in part by the U.S.~Department of Energy through DOE grant DE-SC0009919. WJW is supported by Marie Curie Fellowship PIIF-GA-2012-328913.

\bibliography{lensing}

\end{document}